\documentclass[%
 aip,
rsi,%
 amsmath,amssymb,
 reprint,%
twocolumn
]{revtex4-1}

\usepackage{graphicx}
\usepackage{dcolumn}
\usepackage{bm}
\usepackage{siunitx}
\usepackage{upgreek}
\usepackage{xcolor}
\usepackage[utf8]{inputenc}
\newcommand{\sub}[1]{_{\mathrm{#1}}}
\newcommand{\csub}[1]{$_{\mathrm{#1}}$}
\usepackage[english]{babel}

\usepackage{graphicx}

\usepackage{booktabs}

\DeclareSIUnit\sccm{SCCM}
\DeclareSIUnit\torr{Torr}

\begin{document}

\preprint{AIP/123-QED}

\title{A self-aligned nano-fabrication process for vertical NbN--MgO--NbN\\ Josephson junctions}%

\author{Alexander Grimm}
\affiliation{Université Grenoble Alpes, INAC-PhElIQS, F-38000 Grenoble, France.}%
\affiliation{CEA, INAC-PhElIQS, F-38000 Grenoble, France.}

\author{Salha Jebari}%
\affiliation{Université Grenoble Alpes, INAC-PhElIQS, F-38000 Grenoble, France.}
\affiliation{CEA, INAC-PhElIQS, F-38000 Grenoble, France.}

\author{Dibyendu Hazra}
\affiliation{Université Grenoble Alpes, INAC-PhElIQS, F-38000 Grenoble, France.}
\affiliation{CEA, INAC-PhElIQS, F-38000 Grenoble, France.}

\author{Florian Blanchet}
\affiliation{Université Grenoble Alpes, INAC-PhElIQS, F-38000 Grenoble, France.}
\affiliation{CEA, INAC-PhElIQS, F-38000 Grenoble, France.}

\author{Frédéric Gustavo}
\affiliation{Université Grenoble Alpes, INAC-PhElIQS, F-38000 Grenoble, France.}
\affiliation{CEA, INAC-PhElIQS, F-38000 Grenoble, France.}

\author{Jean-Luc Thomassin}
\affiliation{Université Grenoble Alpes, INAC-PhElIQS, F-38000 Grenoble, France.}
\affiliation{CEA, INAC-PhElIQS, F-38000 Grenoble, France.}

\author{Max Hofheinz}
\affiliation{Université Grenoble Alpes, INAC-PhElIQS, F-38000 Grenoble, France.}
\affiliation{CEA, INAC-PhElIQS, F-38000 Grenoble, France.}

\date{\today}

\begin{abstract}

We present a new process for fabricating vertical NbN--MgO--NbN Josephson junctions using self-aligned silicon nitride spacers. It allows for a  wide range of junction areas from \SI{0.02}{\micro \meter \squared} to several \SI{100}{\micro \meter \squared}. At the same time, it is suited for the implementation of complex microwave circuits with transmission line impedances ranging from $ < \SI{1}{\ohm}$ to $> \SI{1}{\kilo \ohm}$. 
The constituent thin films and the finished junctions are characterized. The latter are shown to have high gap voltages ($> \SI{4}{\milli \volt}$) and low sub-gap leakage currents. 
\end{abstract}


\maketitle

\section{Introduction}
Josephson junctions (JJ) are the key nonlinear element used in
superconducting circuits\cite{Taylor1967,Devoret2013,Clarke2004}. In
applications where large critical currents are needed, typically
Nb--Al$_2$O$_3$--Nb, NbN--MgO--NbN or NbN--TaN--NbN junctions are used. This
is the case for superconducting quantum interference
device (SQUID) magnetometers\cite{Clarke2004}, Josephson
mixers\cite{Tucker1985} or rapid single flux quantum (RSFQ) logic
circuits\cite{Likharev1991}. The high critical temperature
and large superconducting gap of Nb and NbN enable these devices to
operate at liquid helium temperatures (\SI{4}{\kelvin}) and up to very
high frequencies (above \SI{1}{\tera \hertz} for NbN). Such junctions
are typically fabricated by first depositing the junction stack and
then etching a pillar from it. The main difficulty is to contact the
top of this pillar without short-circuiting it to the bottom
layer. Common solutions to this problem, such as an additional
insulating layer that is etched away above the pillar or anodic oxidation of Nb usually limit
the minimum junction area to approximately \SI{1}{\micro \meter
  \squared}~\cite{Nakamura2011, Hypres2012, Hypres2015, Kiviranta2016}. This results 
in high critical currents and large shunting capacitances.

However, certain superconducting quantum circuits, such as Josephson
qubits and Josephson photonic devices, require these quantities to be small. Because of this, it is
much more common to use Al--Al$_2$O$_3$--Al junctions, which can be fabricated
using shadow angle evaporation; a technique allowing for very small
Josephson junction sizes down to \SI{0.01}{\micro \meter \squared} or
less \cite{Dolan1977, Lecocq2011}. The downside of this approach is that the
low gap of Aluminum sets an ultimate limit for the fabricated devices in operation
frequency (below \SI{100}{\giga \hertz}) as well as temperature.

Here, we report on a fabrication process able to simultaneously obtain NbN--MgO--NbN
junctions with surface areas as small as \SI{0.02}{\micro \meter \squared}
and as large as several \SI{100}{\micro \meter \squared}. This marks an important step towards
bringing the advantages of NbN--MgO--NbN junctions to quantum circuits
that were so far limited to Al junctions and opens the possibility for
such circuits to operate at much higher frequencies. It is
particularly useful for the field of Josephson
photonics\cite{Hofheinz2011,Gramich2013,Armour2015,Leppakangas2015,Grimm2015},
which requires small junctions and could provide devices like quantum
microwave sources and amplifiers able to function up to the gap
frequency of the superconductor.

The presented NbN junction fabrication process also enables us to
implement very versatile passive microwave circuits at no extra
cost. The large kinetic inductance of NbN can be used to design high
impedance circuit elements, which are useful for many quantum circuits
in general and Josephson photonics in particular.

\section{Fabrication of Josephson junctions}
\begin{figure}
	\centering
	\includegraphics[width=0.48\textwidth]{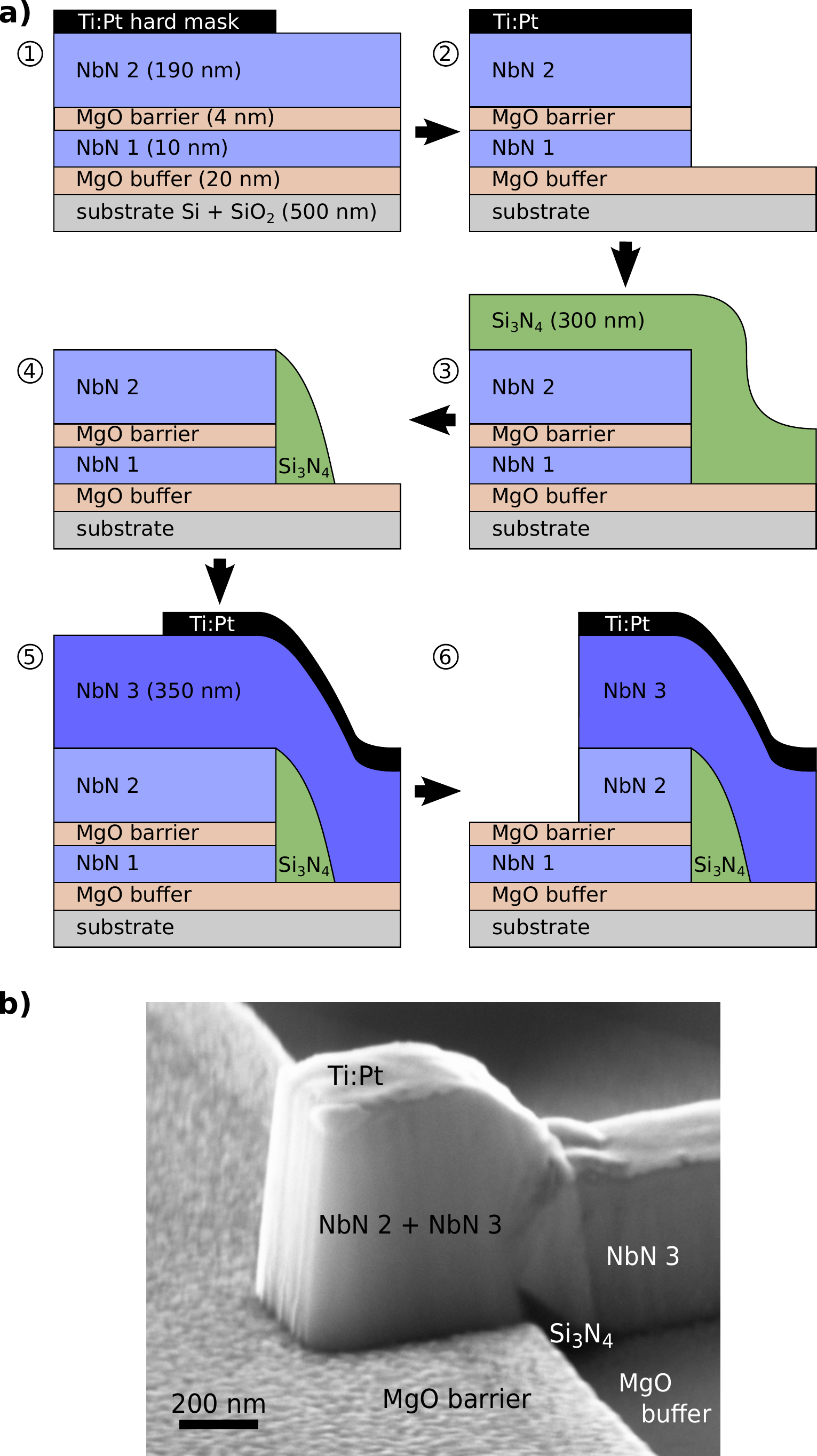}
	\caption{Sample fabrication: (a) The different steps of the fabrication process. First the NbN-MgO-NbN trilayer is sputtered (1) and etched (2) using a Ti:Pt hard mask. Then a dielectric layer is deposited conformally by chemical vapor deposition (3) and etched directionally (using a Ti:Pt hard mask) to leave spacers passivating the trilayer sidewalls (4). Finally a counter electrode layer is deposited (5) and etched down to the junction barrier (6). (b) Finished $\SI{150}{\nano \meter} \times \SI{150}{\nano \meter}$ junction defined by EBL with a Ti:Pt hard mask still in place.}
	\label{fig:fab_process}
\end{figure}

The NbN thin films are prepared by DC magnetron sputtering on
Si($\SI{500}{\micro \meter}$):$\textrm{SiO}_{2}$($\SI{500}{\nano
  \meter}$) substrates. After cleaning the wafers by back-sputtering
in the deposition chamber, a $\SI{20}{\nano \meter}$ magnesium oxide
(MgO) buffer layer is added as etch stop and to improve the superconducting properties
of the NbN film. This is followed by the deposition of an
NbN($\SI{10}{\nano \meter}$):MgO($\SI{4}{\nano
  \meter}$):NbN($\SI{190}{\nano \meter}$) stack (hereafter called the
trilayer) \cite{Larrey1999}. The NbN is DC
magnetron sputtered from an Nb target in an atmosphere of Ar and
N$_2$ (see supplementary information for detailed process parameters). Scanning electron microscope images of cuts through the
produced NbN films show them to be columnar with an average column
diameter of $\approx\SI{20}{\nano \meter}$.

The aspect ratio imposed by the thickness of the deposited layers and the lateral junction sizes calls for very directional etch processes. We have developed several dry etch recipes on an Oxford ICP Plasmalab100 reactive ion etcher. During the etch, NbN is attacked both chemically, with $\textrm{SF}_{6}$, as well as mechanically with Ar. 
A moderate platen power generates an auto-polarization voltage of $\approx\SI{190}{\volt}$ accelerating the ions towards the sample and makes the etch more directional. Moreover, $\textrm{CH}_{2}\textrm{F}_{2}$ is added to the mixture, with the effect of polymerizing the exposed surface of the NbN film, making it less sensitive to the chemical etch. While in the vertical direction the Ar bombardment constantly ablates the polymer film and leaves the surface exposed to the chemical etch, the sidewalls of steps and trenches stay protected, preventing any underetch and assure smooth and steep sidewalls (see Fig.~\ref{fig:fab_process}). 

The MgO-barrier is etched purely mechanically with an argon plasma. MgO is insensitive to the NbN etch described above, allowing us to use the buffer and barrier layers as effective etch stops, despite their very small thickness. Consequently the NbN etch can run longer to counteract inhomogeneities in deposition and etching rate across the wafer.

During the entire fabrication process described below optical lithography (OL) and electron beam lithography (EBL) steps are combined in order to be able to define small structures with high precision, while keeping the processing times for bigger structures such as coplanar waveguide (CPW) transmission lines \cite{Pozar2011} low. The following description will, however, focus on the elaboration of a single small Josephson junction and thus disregards some of the OL steps.

To begin with, a step is etched into the trilayer using an EBL defined Ti($\SI{10}{\nano \meter}$):Pt($\SI{60}{\nano \meter}$) hard-mask (Fig.~\ref{fig:fab_process}), which is subsequently removed with an Ar plasma. Next, the entire wafer is coated with a $\SI{300}{\nano \meter}$ thick film of $\textrm{Si}_{3}\textrm{N}_{4}$ by chemical vapor deposition. A $\SI{10}{\nano \meter}$ thick layer of MgO is deposited on top of the SiN in order to protect it from overetching during some of the following fabrication steps. An optical lithography step is carried out defining regions where the dielectric will be etched. The entire junction lies in such an area.  First, the MgO in these areas is removed by a dip in 1\,\% acetic acid. We then perform a directional dry-etch of the SiN, similar to the one developed for NbN. The vertical thickness of the dielectric at the step is considerably bigger than elsewhere. Hence, it can be etched away on all flat surfaces while leaving behind a self-aligned spacer protecting the side of the trilayer (Fig.~\ref{fig:fab_process}).\\
Finally, after a back-sputtering step, the NbN counter-electrode ($\approx\SI{350}{\nano \meter}$) is deposited. Another EBL defined hard-mask protects an area in the shape of a finger overlapping with the step. Once the unprotected areas are etched down all the way to the MgO tunneling barrier, this finger forms the vertical Josephson junction shown in Fig.~\ref{fig:fab_process} (see supplementary material for details of the etch recipes).

\section{Simultaneous implementation of passive elements}
The fabrication steps used to implement the junction enable us to
simultaneously fabricate very versatile passive microwave circuits on
the same chip. In particular, the process allows for various types of
transmission lines: A line defined in the counter-electrode layer on
top of the SiN forms a microstrip line together with the ground plane
given by the trilayer. This case is ideal for low-impedance
transmission lines (we estimate $\SI{65}{\ohm}$ for \SI{1}{\micro
  \meter} wide lines, down to $< \SI{1}{\ohm}$ for \SI{100}{\micro
  \meter} wide lines. However, these impedances depend critically on
the exact dielectric thickness and, because of high kinetic inductance
fractions around 0.8, on the NbN material properties. Coplanar wave
guide (CPW) geometries using the full thickness of the trilayer can be tuned 
for transmission line impedances between \SI{20}{\ohm} and
\SI{150}{\ohm}~\cite{Grimm2015}. This impedance and propagation speed are accurately
controlled by lateral geometry because kinetic inductance fractions are low ($< 0.25$
except for extreme geometries) and the substrate dielectric constant is well known. For even higher
impedances, the dielectric on top of the CPW center conductor can be
removed before the counter-electrode etch step. As a result, the upper
part of the trilayer is etched away during this step, thinning
down the center conductor to $\SI{10}{\nano \meter}$ like in the
junction area (see Fig.~\ref{fig:fab_process}). For such thin layers
the NbN kinetic inductance is dominating and leads to characteristic
impedances of up to $\SI{3.5}{\kilo\ohm}$ for a $\SI{1}{\micro \meter}$
wide wire.
In areas where the SiN dielectric layer is not etched, it separates the trilayer and the counter-electrode, allowing for straightforward implementation of parallel plate capacitors. Where needed, holes can be etched in the dielectric to provide vias connecting the two layers. This enables us to fabricate crossovers between the ground planes of the CPW transmission lines, thus effectively eliminating parasitic modes~\cite{Abuwasib2013,Chen2014,Kwon2001,Koster1989}.

%
\section{Properties of the fabricated films and junctions}

\begin{figure}
	\centering
	\includegraphics[width=0.5\textwidth]{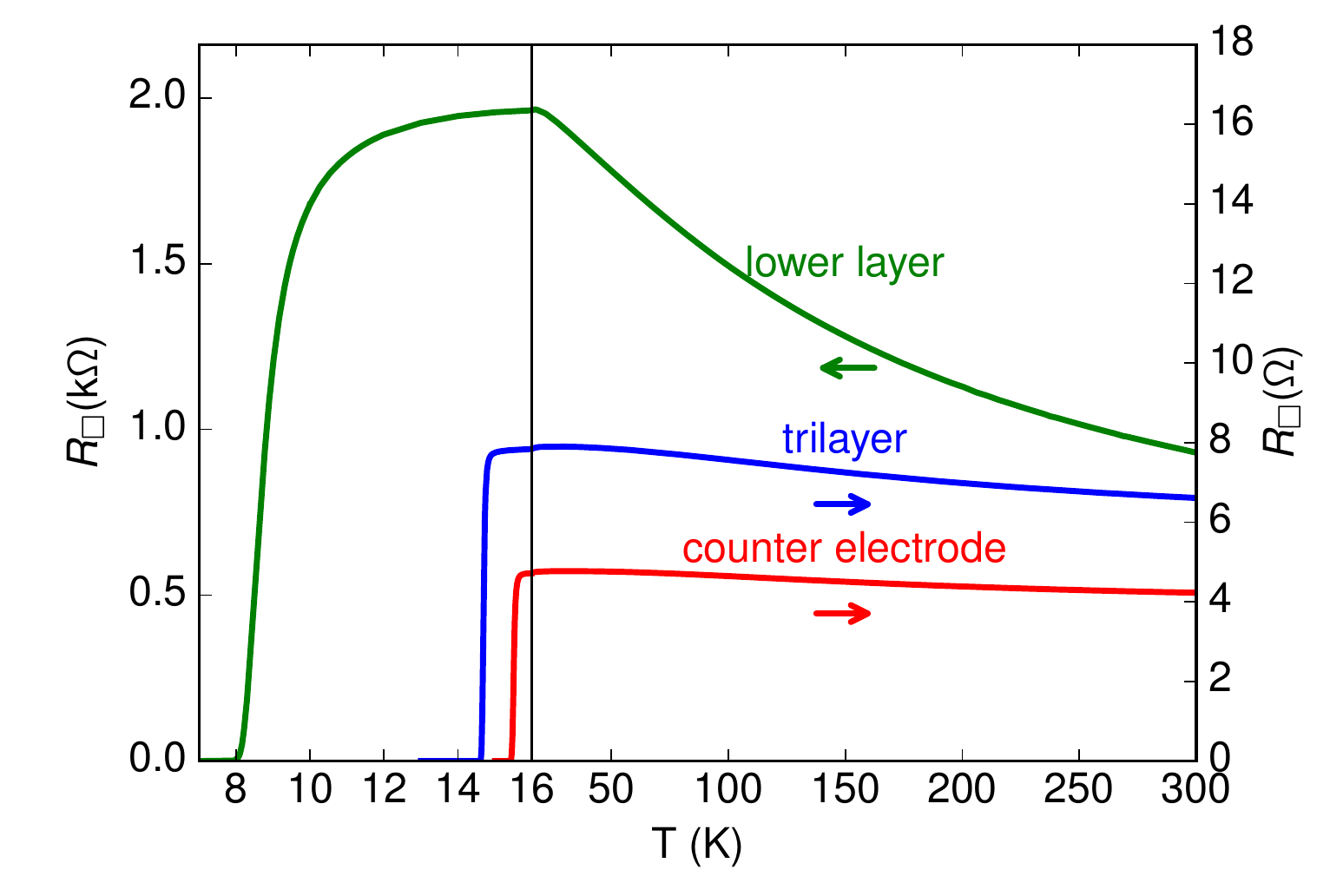}
	\caption{Square resistance as a function of temperature for the different layers of our process. Arrows indicate on which resistance scale the data is plotted. The temperature curves of lower, trilayer and counter electrode layers give a critical temperature of respectively 8.8, 14.7 and $\SI{15.5}{\kelvin}$ and we estimate their surface inductances $L_{\mathrm{kin},\square}$ to be, respectively, 300, 0.81, $\SI{0.55}{\pico \henry}$.}
	\label{fig:fab:Tc}
\end{figure}

Temperature dependent resistance measurements of the deposited layers were performed in a commercial physical property measurement system (PPMS) (see Fig.~\ref{fig:fab:Tc}).

\begin{figure}
	\centering
	\includegraphics[width=0.5\textwidth]{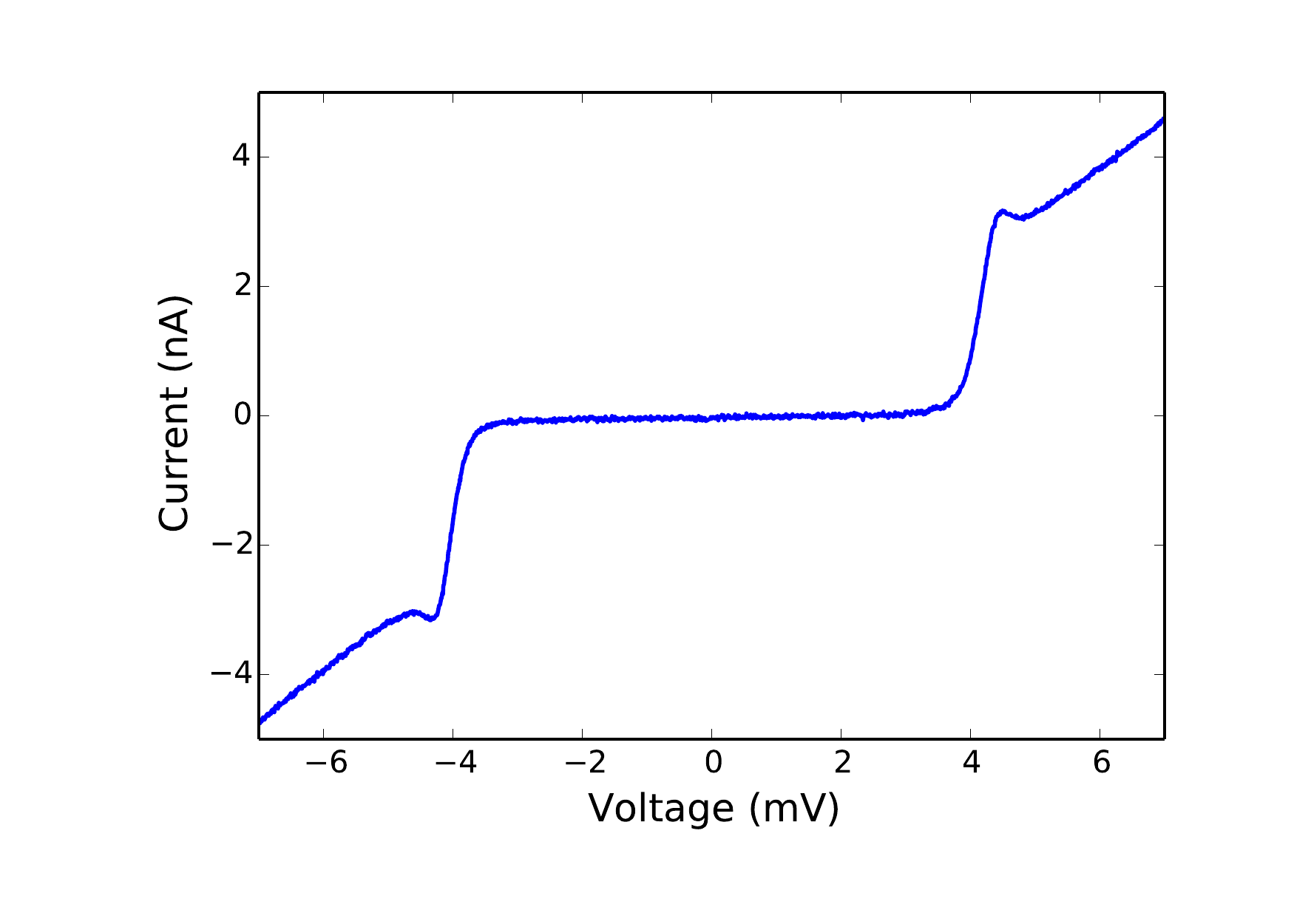}
	\caption{Typical current voltage characteristics. The curve shows the data for a SQUID consisting of two parallel Josephson junctions with dimensions of $\SI{150}{\nano \meter}\times\SI{150}{\nano \meter}$.}
	\label{fig:fab:ivF7B067}
\end{figure}
We calculate the value of the kinetic inductance $L_{\mathrm{kin},\square}$ from the square resistances $R_\square$ (taken at \SI{20}{\kelvin}) of the film, its thickness $d$ and the superconducting gap $\Delta$ estimated from the critical temperature, according to \cite{Kautz1978,VanDuzer1981} $L_{\mathrm{kin},\square}=\mu_{0}\lambda\text{coth}\left(d/\lambda\right)$. Here $\lambda=(\hbar R_\square d/(\mu_{0} \pi \Delta))^{1/2}$ is the London penetration depth of a local superconductor in the low temperature limit $k_{\textrm{B}}T \ll \Delta$\cite{Barends2009,Mattis1958}. For the bottom layer $d \ll \lambda$ so that $L_{\mathrm{kin},\square}\approx\hbar R_\square / (\pi \Delta)$ can be estimated without knowing the exact film thickness.

A typical current-voltage characteristic of a superconducting interference device (SQUID) consisting of two parallel junctions with a total area of $\SI{0.04}{\micro \meter \squared}$ (squares of $\SI{150}{\nano \meter}$ side-length) measured at \SI{4.2}{\kelvin} is shown in Fig.~\ref{fig:fab:ivF7B067}. The gap voltage $V\sub{gap}$ can be defined as the voltage at the steepest part of the curve, where $\textrm{d}I/\textrm{d}V$ is maximized \cite{Kawakami2001}. Here, $V\sub{gap}=\SI{4.15}{\milli \volt}$ corresponding to an emission frequency of $\approx \SI{1}{\tera \hertz}$ in the framework of Josephson photonics \cite{Hofheinz2011,Ingold2005,Grimm2015}. A normal state resistance $R\sub{N}\approx\SI{1.5}{\mega \ohm}$ can be extracted from the same curve far above $V\sub{gap}$ at $\SI{7}{\milli \volt}$. Together with the Ambegaokar-Baratoff formula at zero temperature\cite{Tinkham2004} $I\sub{c}R\sub{N}=\pi\Delta(0)/2e$ and the relation $2\Delta=eV\sub{gap}$ we can evaluate the theoretical critical current at zero temperature to be $\approx\SI{2.2}{\nano \ampere}$.
The critical current density is $\approx\SI{5.5}{\ampere \per \centi \meter \squared}$. Note that, even though our junctions are optimized for low critical current densities, $J\sub{c}$ can easily be increased by several orders of magnitude with this deposition process \cite{Larrey1999}. The current branch of the IV in~Fig.~\ref{fig:fab:ivF7B067} is not visible, because the associated energy scale\cite{Tinkham2004} $E\sub{J}=\Phi_{0}I\sub{c}/(2 \pi)\approx\SI{4.5}{\micro \electronvolt}$, where $\Phi_{0}$ is the magnetic flux quantum, is much smaller than the thermal energy at \SI{4.2}{\kelvin} ($k\sub{B}T\approx\SI{360}{\micro \electronvolt}$). The "knee" in the current-voltage characteristic is usually associated with the formation of a thin normal metal layer close to the junction barrier and has frequently been observed in NbN--MgO--NbN Josephson junctions\cite{Shoji1987,Shoji1991,Longobardi2011}.\\
In the context of this work we took particular interest in minimizing the leakage current under the gap \cite{Grimm2015}. In order to quantify the subgap leakage we compare the resistance under the gap ($R\sub{S}$) at $\SI{3}{\milli \volt}$ to $R\sub{N}$. A linear fit of the current voltage characteristic in Fig~\ref{fig:fab:ivF7B067} between $\SI{-3}{\milli \volt}$ and $\SI{3}{\milli \volt}$ gives $R\sub{S}\approx \SI{80}{\mega \ohm}$ leading to $R\sub{N}/R\sub{S}<0.02$. While the measured gap voltage on this specific sample is lower than the best values reported in the literature, the subgap resistance is comparable to the numbers achieved in other groups \cite{Kawakami2001,Longobardi2011,Shoji1985}. In particular, it should be pointed out that the smallest NbN--MgO--NbN tunnel junctions found in the literature have surface areas around $\SI{0.1}{\micro \meter \squared}$, several times bigger than the junction presented here, while showing considerably more subgap leakage\cite{Kawakami2001,Hunt1989,LeDuc1991}.

\section{Possible variations}
The current process is not suitable for applications requiring resonators with very high quality factors, such as circuit QED. CPW resonators realized with this process\cite{Grimm2015} typically have intrinsic quality factors of $10^4$. We attribute this low value to dielectric loss in the MgO and $\textrm{SiO}_{2}$ buffer layers. We expect a significant increase of this value by omitting the MgO buffer layers and using sapphire substrates instead. The latter allow to directly grow high quality NbN films without buffer layer and provide a good etch stop. If transmission lines with higher impedances are needed, the thickness of the lower NbN layer can be reduced. Replacing the silicon substrate with a silica substrate, increases the characteristic impedances of CPWs by another factor $\sim 1.7$.

If high kinetic inductance is not desired, the process, with slight adjustments, should be applicable also to Nb-(Al)-Al$_2$O$_3$-Nb trilayers, where the barrier is formed by oxidizing a thin proximized aluminum layer. We expect that such a process yields lower leakage current below the gap because of the more uniform self-passivating barrier. 

The current density is voluntarily kept low for the application in mind \cite{Grimm2015}, but can be increased by several orders of magnitude by decreasing the thickness of the MgO barrier layer \cite{Larrey1999}.

Finally, if only larger junction sizes are needed, the junctions can exclusively be defined by optical lithography, considerably reducing fabrication times. In this case, the geometry is changed so that a counter-electrode wire crosses a trilayer wire. This makes the junction size insensitive to slight misalignments between the two optical lithography steps.

\section{Conclusion}
In conclusion, we have developed a new fabrication process for vertical NbN--MgO--NbN Josephson  with self-aligned SiN spacers allowing for very small junction areas. The measured junction current--voltage characteristics show  a high gap voltage and to our knowledge the lowest subgap leakage current at these junction sizes reported so far. The process allows for simultaneous fabrication of very large Josephson junctions and extremely versatile passive microwave circuits with characteristic impedances ranging from $< \SI{1}{\ohm}$ to $> \SI{1}{\kilo \ohm}$ .

\begin{acknowledgments}
We acknowledge financial support from the Grenoble Nanosciences Foundation grant JoQOLaT, from French National Research Agency / grant ANR14-CE26-0007 - WASI,  and from the European Research Council under the European Unions Seventh Framework Programme (FP7/2007-2013) / ERC Grant agreement No 278203 WiQOJo, as well as fruitful discussions with the Plateforme Technologique Amont (PTA) cleanroom team, François Lefloch and Jean-Claude Villegier.
\end{acknowledgments}

\section*{Supplementary material: Detailed process parameters}
  \subsection{Film deposition}
NbN and MgO films are deposited in an Alcatel SCM600 sputter tool. The NbN--MgO--NbN trilayer is deposited \emph{in situ} without exposing the sample to air between steps. Substrates are not actively heated, but their temperature increases above ambient temperature during the process.

NbN films prepared by reactive DC magnetron sputtered from a \SI{150}{\milli \meter} Niobium target at \SI{4}{\ampere} and $\approx \SI{400}{\volt}$ in an atmosphere of \SI{1.5}{\pascal} of Ar and \SI{0.2}{\pascal} of N$_2$. The latter partially nitrates the target surface after the target surface has been conditioned in an atmosphere containing more nitrogen. The deposition rate is $\approx\SI{3.3}{\nano \meter \per \second}$. For very thin layers we rotate the sample stage over different targets at $\approx\SI{0.6}{\hertz}$ reducing the average deposition rate by approximately a factor 7 for a more precise control of thickness.

The MgO barrier is RF magnetron sputtered at \SI{450}{\watt} in an Ar atmosphere at \SI{1.3}{\pascal} for \SI{100}{s},  in the same chamber as the NbN films, also with rotating sample. For deposition of the buffer layer we increase power to \SI{550}{\watt}, keep the sample on top of the target and use an atmosphere of \SI{1.25}{\pascal} of Ar and \SI{0.1}{\pascal} of N\csub{2}.

SI\csub{3}N\csub{4} is deposited in a Corial D250L plasma enhanced chemical vapor deposition (PECVD) tool at \SI{280}{\celsius} and \SI{200}{\watt} at a pressure of \SI{200}{\pascal} with flows of \SI{100}{\sccm} of SiH\csub{4}, \SI{500}{\sccm} of NH\csub{3} and \SI{100}{\sccm} of Ar.

\subsection{Etch parameters}
Tables \ref{tab:etch-parameters_tri}, \ref{tab:etch-parameters_sin}, \ref{tab:etch-parameters_resist} give details on our etch recipes in an Oxford Plasmalab 100 ICP etcher. The different rows of each table correspond to the steps followed chronologically during the procedure. Landing steps are performed to homogenize the NbN etch over the entire wafer and are possible because MgO provides a very good etch stop. Values in parentheses are used to start the plasma and then quickly ramped down to the regular parameters. The purely mechanical MgO etch steps are interleaved with pumping and venting of the process chamber in order to evacuate excess material. EPD stand for ``end point detection'', which is performed using \textit{in situ} laser reflectometry.

\begin{table*}[ht]
\begin{tabular}{| c | l | c | c | c| c | c | c | c |}
\hline
\multicolumn{2}{|l|}{} & {CH\csub{2}F\csub{2} (\si{\sccm})} & {SF\csub{6} (\si{\sccm})} & {Ar (\si{\sccm})} & {Pressure (\si{\milli\torr})} & {ICP (\si{\watt})} & {Forward (\si{\watt})} & {Time (\si{\second})} \\
\hline      
\multicolumn{2}{|l|}{upper NbN} & 25 & 5 & 40 & 5(15) & 500 & 70 & {EPD} \\ \hline
\multicolumn{2}{|l|}{NbN landing} & 10 & 10 & 40 & 20 & 500 & 20 & 15 \\ \hline 
   &MgO & &  & 100 & 5 & 500 & 150 & 20 \\ \cline{2-9}
$4\times$ &Pump &  &  &  & 0  &  &  & 150 \\ \cline{2-9}
   &Vent &  &  &  & {atmospheric pressure} &  &  & 90 \\ \hline
\multicolumn{2}{|l|}{lower NbN} & 25 & 5 & 40 & 5(15) & 500 & 70 & {EPD} \\ \hline
\multicolumn{2}{|l|}{NbN landing} & 10 & 10 & 40 & 20 & 500 & 20 & 15 \\
\hline
\end{tabular}
\caption{Procedure of etching the trilayer and counter electrode. Thin lower layers (below approximately $\SI{30}{\nano \meter}$) are transparent and allow for end point detection in the first step. If thicker lower layers are required, the first etch has to be timed instead. For the final etch of the counter electrode (and upper part of the trilayer, see paper) the etch is stopped before the MgO step.}
\label{tab:etch-parameters_tri}
\end{table*}

\begin{table*}[ht]
\centering
\begin{tabular}{| c | c | c | c| c | c | c | c |}
\hline & CH\csub{2}F\csub{2} (\si{\sccm}) & SF\csub{6}  (\si{\sccm})& Ar (\si{\sccm}) & Pressure (\si{\milli\torr}) & ICP (\si{\watt}) & Forward (\si{\watt})& Time (\si{\second}) \\ \hline
Si\csub{3}N\csub{4} & 25 & 5 & 40 & 5(15) & (1000,500),250 & (150),50 & EPD \\ \hline
\end{tabular}
\caption{Procedure for etching Si\csub{3}N\csub{4} dielectric. Before this etch the MgO cap layer is removed by dipping the wafer in 1\,\% acetic acid for $\SI{20}{\second}$.}
\label{tab:etch-parameters_sin}
\end{table*}

\begin{table*}[ht!]
\centering
\begin{tabular}{| c | c | c | c| c | c | c | c |}
\hline & O\csub{2} (\si{\sccm})&  Pressure (\si{\milli\torr}) & ICP (\si{\watt}) & Forward (\si{\watt})& Time (\si{\second})\\ \hline
O\csub{2} plasma & 45 & 30 & 500 & 0 & 300 \\ \hline
\end{tabular}
\caption{Etch for removing resist residues after etching and stripping.}
\label{tab:etch-parameters_resist}

\end{table*}


%

\end{document}